\documentclass[preprint,3p,twocolumn]{elsarticle}
\usepackage{geometry}
\usepackage{fancyhdr}

\biboptions{numbers,sort&compress}
\usepackage{graphicx}
\usepackage{amsmath}
\usepackage{amssymb}
\usepackage{ulem}
\usepackage{url}
\usepackage{nidanfloat}

\begin{document}

\begin{frontmatter}

\title{Comparing mesoscopic models for dendritic growth}

\author{Damien Tourret$^1$, Laszlo Sturz$^2$, Alexandre Viardin$^2$, Miha Zalo\v znik$^3$}
\address{
$^1$\,IMDEA Materials Institute, Getafe, Madrid, Spain\\
$^2$\,Access e.V., Aachen, Germany\\
$^3$\,Universit\'e de Lorraine, CNRS, IJL, F-54000 Nancy, France
}
\address{\rm\url{damien.tourret@imdea.org}}

\begin{abstract}

We present a quantitative benchmark of multiscale models for dendritic growth simulations. We focus on approaches based on phase-field, dendritic needle network, and grain envelope dynamics. As a first step, we focus on isothermal growth of an equiaxed grain in a supersaturated liquid in three dimensions. 
A quantitative phase-field formulation for solidification of a dilute binary alloy is used as the reference benchmark. 
We study the effect of numerical and modeling parameters in both needle-based and envelope-based approaches, in terms of their capacity to quantitatively reproduce phase-field reference results. 
In light of this benchmark, we discuss the capabilities and limitations of each approach in quantitatively and efficiently predicting transient and steady states of dendritic growth. 
We identify parameters that yield a good compromise between accuracy and computational efficiency in both needle-based and envelope-based models.
We expect that these results will guide further developments and utilization of these models, and ultimately pave the way to a quantitative bridging of the dendrite tip scale with that of entire experiments and solidification processes.

\end{abstract}

\end{frontmatter}
\thispagestyle{fancy}

\section{Introduction}

In metallic alloys obtained by solidification processing, dendritic microstructures are common \cite{flemings}.
From a fundamental standpoint, dendritic growth theory and modeling stands as a challenge to combine phenomena across a wide range of scales, from microscopic capillarity at dendritic tips to macroscopic transport of heat and solute in the melt \cite{Langer1980,Trivedi1994}.
Due to this multiscale aspect, numerous approaches have emerged over the years that aim at bridging length scales in solidification modeling \cite{boettinger2000solidification,Asta2009a,karma2016atomistic}.
Each of these approaches operates within a different range of scales, which makes them valuable tools from a technological innovation perspective, since a key hurdle on the way to effective ICME (Integrated Computational Materials Engineering) implementations relies upon our ability to couple models at different length scales \cite{allison2011integrated, panchal2013key}.
Yet, most modeling approaches to dendritic growth have thus far been developed separately, with little effort to compare them and to discuss their capabilities and limitations on a quantitative basis.

In this article, we focus on models of solidification that operate from the microscopic to an intermediate, or {\it mesoscopic}, scale between that of the {\it microscopic} solid-liquid interface pattern and that of the {\it macroscopic} solidification process.
At this intermediate scale, interactions between grains occur that determine microstructural features such as grain size, shape, morphology, and crystal texture.
Mesoscopic models are more recent than their microscopic and macroscopic counterparts, such that a rigorous quantitative assessment of their relative advantages and limitations remains to be performed.
We also focus on moderate solute supersaturation, $\Omega\leq 0.25$, which is a common regime for most processes that do not involve rapid solidification.

Here, we focus on three prominent models for dendritic crystal growth --- namely phase-field, dendritic needle network, and mesoscopic grain envelope models. 
Using a set of benchmark simulations for isothermal equiaxed growth in a binary alloy at given undercoolings (i.e. at different solute supersaturations), we compare the predicted growth velocities of the dendrite tips in both steady and transient growth regimes.
The objective of this comparison is to highlight the advantages and limitations of the different methods, and to suggest how to select model parameters in order to achieve quantitative predictions while retaining profitable computational efficiency.
While this study is still ongoing, the preliminary results presented here already provide useful insight into the capabilities of the models and the choice of parameters.

\section{Models}

\begin{figure*}[b]
\centering
 \includegraphics[width=.78\textwidth]{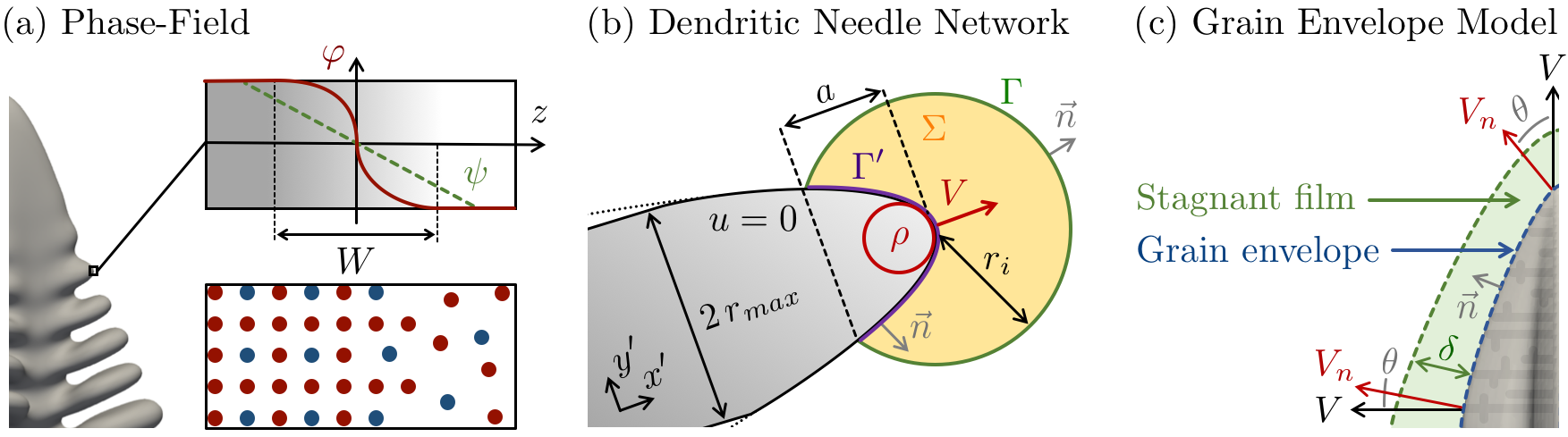}
\caption{
Schematic illustration of the phase field (a), dendritic needle network (b), and envelope (c) models, with characteristic model parameters, namely the PF diffuse interface width $W$, the DNN integration radius $r_{i}$ and bounding radius $r_{max}$, and the stagnant-film thickness $\delta$.}
\label{fig:schematics}
\end{figure*}

\subsection{Phase-field}
\label{sec:phasefield}

The phase-field (PF) method is a powerful tool for the resolution of free-boundary problems.
The method can handle moving interfaces of arbitrary morphological complexity.
Interfaces and boundaries are not tracked explicitly, but instead different phases or grains are represented by a ``phase field'' that varies continuously through an interface of finite width $W$ (Figure~\ref{fig:schematics}a). 

The free energy of the system is formulated, typically using a Ginzburg-Landau functional~\cite{ginzbuglandau,halperin1974renormalization}, which describes the energetic contributions of interfaces, transformations, and local state (e.g. temperature, concentration, stresses, etc.)~\cite{chen2002phase,boettinger2002phase,Moelans2008,steinbach2009phase,Plapp2016,tonks2019phase}.
The time evolution of the fields derives directly from the functional derivatives of the constructed free energy functional, using the so-called Cahn-Hilliard equation \cite{cahn1958free} for a conserved field, such as the solute concentration, or a time-dependent Ginzburg-Landau (also referred to as Allen-Cahn \cite{allen1979microscopic}) equation for non-conserved fields.
A number of PF models have been proposed, including some that consider a grand potential functional in place of a free energy description \cite{Plapp2011,choudhury2012grand}, but the basic recipe typically follows these same common traits for every phase-field model.
Hence, what makes the distinction between different PF models is essentially the choice of the number of phase fields (or order parameters) necessary to describe the system and the transformation(s), the mathematical/physical formulation of the functional, the interpolation functions used for different properties between different phases, and the matching of the model parameters to physical parameters relevant to the phenomena one aims to describe.

In the context of solidification, the PF method has been particularly successful, in part due to the presence of well-defined sharp interface problems to be matched by the PF formulations~\cite{boettinger2002phase,steinbach2009phase}. 
In particular, ``thin-interface'' asymptotic analyses have permitted using a diffuse interface width $W$ much higher than the actual width of the solid-liquid interface \cite{karma1998quantitative,Echebarria2004}.
The introduction of a so-called ``anti-trapping current'' has allowed quantitative simulations of alloys by eliminating spurious numerical solute trapping for wide interfaces \cite{Karma2001a,Echebarria2004}. 
In the model considered here, the temperature is assumed homogeneous and constant in time, solid-state diffusion and kinetic undercooling are neglected, a binary alloy is considered within its dilute limit where its phase diagram can be linearized with a constant interface solute partition coefficient $k$, and solute transport in the liquid is limited to diffusion with a constant diffusivity $D$.
These assumptions result in the following evolution equations \cite{karma1998quantitative,Karma2001a,Echebarria2004,Tourret2015}
\begin{align} 
\label{eq:psi}
a_s({\mathbf n})^2 \frac{\partial\psi}{\partial t} = 
\nabla \cdot \left[a_s({\mathbf n})^2 \nabla\psi \right] 
 + \varphi \sqrt 2 
 - \lambda \sqrt 2 (1-\varphi^2) U	
\nonumber\\
 + \sum_{m=x,y,z}\left[ \partial_m\left( |\nabla\psi|^2 a_s({\mathbf n}) \frac{\partial a_s({\mathbf n})}{\partial(\partial_m\psi)} \right) \right] \\
\label{eq:u}
\Big( 1+k-(1-k)\varphi \Big) \frac{\partial U}{\partial t} = 
\widetilde{D} \; \vec\nabla\cdot\left[ (1-\varphi)\vec\nabla U\right]  
\nonumber\\
 +\Big[ 1+(1-k)U \Big] \frac{(1-\varphi^2)}{\sqrt{2}} \frac{\partial\psi}{\partial t} \nonumber\\
 +\vec\nabla\cdot\left[ \Big(1+(1-k)U\Big) \frac{(1-\varphi^2)}{2} \frac{\partial\psi}{\partial t} \frac{\vec\nabla\psi}{|\vec\nabla\psi|} \right] ,
\end{align}
where the dimensionless solute field is
\begin{equation} 
U = \frac{1}{1-k}\left[ \frac{c/c_l^0}{(1-\varphi)/2 + k(1+\varphi)/2 }-1\right] 
\end{equation}
with $c$ the solute concentration field, $c_l^0$ the equilibrium solute concentration of a flat interface at the considered temperature $T_0$.
In these equations, space is scaled with the diffuse interface width $W$ and time with the relaxation time $\tau_0$ at $T_0$. Non-dimensional values of the liquid diffusion coefficient $\widetilde{D}=D\tau_0/W^2=a_1a_2W/d_0$ and the coupling factor $\lambda=a_1W/d_0$, with $a_1=5\sqrt2/8$, and $a_2=47/75$ from thin-interface asymptotic analysis~\cite{karma1998quantitative}.
We use the standard cubic symmetry of the surface tension with $a_s(\mathbf{n}) = (1-3\epsilon_4) + 4\epsilon_4(n_x^4+n_y^4+n_z^4)$, where $n_x$, $n_y$, and $n_z$ are the components of the unit vector $\mathbf{n}$ normal to the interface, and $\epsilon_4$ is the strength of the surface tension anisotropy. 
The classical phase field $\phi$, with $\phi=+1$ in the solid and $\phi=-1$ in the liquid phase, was replaced by a preconditioned phase field $\psi$, defined by $\phi=\tanh(\psi/\sqrt{2})$. 
This change of variable stabilizes the numerical resolution of the equations, thus allowing the use of a coarser grid spacing with negligible loss in accuracy~\cite{glasner2001nonlinear}.
The model is solved using finite differences on a homogeneous grid of cubic elements, and an explicit Euler time stepping scheme.
The code is written in cuda (Compute Unified Device Architecture) programming language, which allows its acceleration through multi-threading on Nvidia graphics processing units (GPUs).

\subsection{Dendritic Needle Network}
\label{sec:dnn}

The DNN approach aims at bridging the scale of the dendrite tip radius $\rho$ and the larger scale of transport phenomena, e.g. the solutal diffusion length $l_D=D/V$, with $V$ the crystal growth velocity. 
The model, rigorously derived in the limit of small P\'eclet numbers, $P=\rho/(2l_D)\ll 1$, represents a dendritic crystal as a network of thin needles that approximate the primary stems and higher order side-branches of the grain. 
In earlier descriptions of the model, in two dimensions, the needle-like branches were treated as infinitely sharp \cite{Tourret2013}.
In later versions of the model, branches were given a parabolic shape, which lead to the same governing equations, while also allowing a three-dimensional extension of the model \cite{Tourret2016a}.

The DNN model was shown to quantitatively reproduce primary dendritic spacings measured in directional solidification experiments in different Al-based alloys \cite{tourret2015threea,tourret2015threeb,Tourret2016a}. Implementations have been developed that include fluid flow in the liquid \cite{tourret2019multiscale,isensee2020} and the interaction between nucleation and growth~\cite{Sturz2016,geslin2016numerical}.

The evolution of the network during solidification is described by the growth dynamics of each needle tip and the generation of new sidebranches.
For a binary alloy, typical assumptions of the model include a linear phase diagram with an interface partition coefficient $k$, negligible fluid flow in the liquid phase (at least in the vicinity of the dendrite tips~\cite{tourret2019multiscale}), no solute transport in the solid phase, and negligible release of latent heat during solidification. 
Under these conditions, for an isothermal domain at a given solute supersaturation $\Omega$, in three dimensions, the needle tip radius, $\rho$, and its velocity, $V$, are given by combining the following set of equations \cite{Tourret2016a}: a solvability condition at the needle tip scale $\approx\rho$
\begin{align}
\label{eq:r2v}
\rho^2V=2Dd_0/\sigma ,
\end{align}
a solute conservation at intermediate scale much larger than $\rho$ but much smaller than $l_D$
\begin{equation}
\label{eq:rv}
\rho V =D \cal{F} ,
\end{equation}
and solute diffusion in the bulk liquid at a scale $\approx l_D$
\begin{equation}
\partial_t u =D \nabla^2 u ,
\end{equation}
with 
\begin{equation}
u=\frac{c_l^0-c}{(1-k)c_l^0}
\end{equation}
the normalized solute concentration, $c_l^0$ the liquid equilibrium concentration and $d_0$ the chemical capillary length, both at the fixed temperature $T_0$, and $\sigma$ the tip selection parameter. 
The flux intensity factor $\cal F$ is a measure of the solute flux towards the needle in the vicinity of its tip up to a distance $a$ behind the needle tip (Figure~\ref{fig:schematics}b), and captures intra- and intergranular solutal interaction. 
It is defined as
\begin{align}
\label{eq:fif}
{\cal F} =~& 
\frac{1}{2\pi a} \; \iint_{\Gamma'}\frac{\partial u}{\partial n}{\rm d}\Gamma' 
\nonumber\\
\approx~& \frac{1}{2\pi a} \; \left( \iint_{\Gamma}\frac{\partial u}{\partial n}{\rm d}\Gamma' + \frac{V}{D} \iiint_{\Sigma}\frac{\partial u}{\partial x'}{\rm d}\Sigma 
 \right),
\end{align}
where $a$ is a chosen integration distance behind the tip, typically a few tip radii, and $\Gamma'$ is the resulting surface of the paraboloid up to a distance $a$ from the tip (Figure~1b). In Eq.~\eqref{eq:fif}, the last equality assumes that the field $u$ is Laplacian in a moving frame of velocity $V$, i.e. $D\nabla^2 u = -V \partial u/\partial x'$, at the scale of the integration domain, with $x'$ the principal growth direction of the branch. 
This assumption allows calculating the normal solute flux along the surface $\Gamma'$ using the normal solute flux through any surface $\Gamma$ that intersects the parabolic tip at a length $a$ behind the tip, with an enclosed volume $\Sigma$ (Figure~\ref{fig:schematics}b)~\cite{Tourret2016a}.
In earlier model implementations \cite{Tourret2013,Tourret2016a,tourret2015threea,tourret2015threeb}, the outer surface $\Gamma$ was set as a square or a cuboid for their ease of implementation onto a square numerical grid. 
In later works, a circular or spherical shape of the integration domain is preferred \cite{tourret2019multiscale,isensee2020}, which facilitates the simulation of grains of arbitrary orientations regardless of the numerical grid.

With $c_\infty$ the alloy nominal solute concentration, the supersaturation $\Omega=(c_l^0-c_\infty)/[(1-k)c_l^0]$ acts as boundary condition far from the interface, where $u\to\Omega$, while thermodynamic equilibrium neglecting capillary correction is applied along the dendritic branches, with $u=0$.

We solve the equations numerically on a 3D finite difference mesh with a homogeneous spacing $\Delta x$, and an explicit time stepping scheme with a constant time step $\Delta t$. 
At each time step, we integrate the value of ${\cal F}(t)$ using Eq.~\eqref{eq:fif} with a spherical integration domain of radius $r_{i}$ centered on the tip of the needle~\cite{tourret2019multiscale}, which provides the value of the tip velocity $V(t)$ and radius $\rho(t)$ from Eqs~\eqref{eq:r2v} and \eqref{eq:rv}.

Far behind its tip, we bound the maximum thickness of a branch to a cylinder of radius $r_{max}$ (Figure~\ref{fig:schematics}b).  
While it does not affect the paraboloid shape in the tip vicinity, this allows a tip to slow down and stop with $V\to0$ without numerical instabilities that would otherwise be induced by the constancy of $\rho^2V$.
Sidebranches normal to the growth direction (here assuming $\langle100\rangle$ growth directions) are generated at positions $\delta=(N_{br}+\delta_{br})\times\rho_s$ behind the tip in all possible directions, when the tip has advanced by a distance $N_{br}\times\rho_s$ since the last branching event. 
A random fluctuation $\delta_{br}$ is generated for every branching event within a range of amplitude $\Delta_{br}$ .
The average distance between sidebranches $N_{br}$ and its fluctuation $\Delta_{br}$, both expressed in units of the theoretical steady-state tip radius $\rho_s$, are user input parameters.
Both truncating and branching methods are similar to those presented and discussed in previous works~\cite{Tourret2013,Tourret2016a,tourret2015threea,tourret2015threeb,tourret2019multiscale,isensee2020,Sturz2016}.

\subsection{Mesoscopic Grain Envelope Method}
\label{sec:envelope}

The Mesoscopic Grain Envelope Model (GEM)~\cite{Steinbach1999,Steinbach2005} bridges the scales from the dendrite tip, over the internal grain structure, to the mesoscopic scale of an ensemble of grains that interact by diffusion and convection of solute. The core idea of the GEM is that solutal interactions at the mesoscopic scale can be accurately simulated without a detailed representation of the ramified structure of the solid-liquid interface. Instead, the description of a dendritic grain can be simplified and represented by its envelope, an artificial smooth surface that links the tips of the actively growing dendrite branches (Figure~\ref{fig:schematics}c). Individual branches are not represented by the model, although, by definition, the growth of the envelope depends on the growth speed of the branch tips. The growth of the tips is controlled by the solute flux that they reject into the liquid and is therefore determined by the local solute concentration of the surrounding liquid. In the envelope model a constitutive model of tip kinetics is used that gives the tip growth speed as a function of the solute concentration field adjacent to the envelope. 

The model was shown to be capable of quantitatively predicting grain envelope shapes, growth velocities, and internal solid fraction for a single grain growing into an infinite melt~\cite{Steinbach1999,Souhar_CMS_2016,Viardin_AM_2017} and for transient interactions among several grains~\cite{Steinbach2005, Olmedilla_AM_2019}. 
The envelope model has been successfully used for simulations of {\it in situ} synchrotron X-ray imaging experiments~\cite{Delaleau2010, Olmedilla2019a, Olmedilla_AM_2019}, 
for modeling spacing adjustments and growth competition in columnar growth~\cite{Viardin_AM_2017,Li_ICASP_2019},
and for characterizing grain shapes and growth kinetics under the influence of solutal interactions used in upscaling to macroscopic models~\cite{TorabiRad_Mtl_2019}. The model was also extended to include fluid flow and solute advection~\cite{Delaleau2011, Olmedilla2019a}.

Two key assumptions are made: (i) that the phenomena controlling the growth of a dendrite tip are universal and are therefore valid for tips of any order (primary, secondary, tertiary,\ldots) and (ii) that the characteristic time of the interactions at the small (tip) scale is much smaller than that of the interactions at the large (grain) scale. In this work we further assume an isothermal domain at a given solute supersaturation, $\Omega$, in three dimensions, and we formulate the model as follows. The normalized concentration at the tip interface is that of thermodynamic equilibirium, neglecting capillary corrections: $u=0$. A stagnant-film formulation of the Ivantsov solution is written that relates the normalized concentration difference, $ u_\delta$, between the liquid at the tip and that at a finite distance, $\delta$, from the tip~\cite{Cantor1977} to the growth P\'eclet number of the tip, $P=\rho V/(2D)$, as
\begin{equation}
\label{eq:cantorvogel}
u_\delta = P \exp(P) \left\{ {\rm E}_1(P) - {\rm E}_1\left( P \left[ 1+\frac{2\delta}{\rho} \right]\right) \right\} .
\end{equation}
Eq.~\eqref{eq:cantorvogel} is used to solve for $P$. The concentration $u_\delta$ is obtained from the concentration field in the liquid around the envelope, which is resolved numerically as explained below. 
The tip selection criterion $\rho^2V=2Dd_0/\sigma$, is a supplementary relation that gives the tip radius, $\rho = d_0/(\sigma P)$, needed in Eq.~\eqref{eq:cantorvogel}, and the tip speed, $V = 2\sigma P^2 D/d_0$. 
Dendrite tips are assumed to grow in predefined growth directions.
In the present study, the cubic crystal dendrites are given six possible growth directions perpendicular to one another. 
The normal envelope growth velocity, $V_{n}$, is calculated from the local tip speed, $V$, by the relation $\vec V_{n}=V\vec n\cos(\theta)$, where $\theta$ is the angle between the outward normal to the envelope, $\vec n$, and the preferential growth direction that forms the smallest angle with $\vec n$ (Figure~\ref{fig:schematics}c). To track the envelope on a numerical grid we use a phase-field-like interface tracking method~\cite{Sun2007}, combined with a surface reconstruction method for better accuracy~\cite{Souhar_CMS_2016}.
Combining these methods, one can avoid explicit tracking of the envelope, which is defined throughout the whole domain with a level set (iso-surface) of a continuous indicator field. Details are given in references \cite{Souhar_CMS_2016,Sun2007}.

Solute transport at the mesoscopic scale, here by diffusion, is described by volume-averaged equations that are valid in the whole domain, both inside and outside the envelopes. 
Solidification inside the envelope uses the Gulliver-Scheil assumptions of thermodynamic equilibrium at the solid-liquid interface, negligible diffusion in the solid, and instantaneous diffusion in the liquid \cite{Scheil_ZM_1942}. 
The resulting liquid fraction field, $g_l$, in the interior of the envelopes, was shown to realistically represent the distribution of solid and liquid within an actual dendritic grain envelope~\cite{Steinbach1999,Souhar_CMS_2016,Viardin_AM_2017}. 
The liquid inside the envelope is at the equilibrium concentration, $u_l=0$. These assumptions lead to the conservation equation for the solute in the liquid phase
\begin{equation}
\label{eq:solute}
g_l \partial_t u_l = D \nabla\cdot\left( g_l\nabla u_l\right) + \partial_t g_l .
\end{equation}
The solution of Eq.~\eqref{eq:solute} gives the liquid concentration $u_l$ outside the envelope and $g_l$ inside the envelope. Outside the envelope, $g_l=1$ and Eq.~\eqref{eq:solute} reduces to a single phase diffusion equation. 
Inside the envelope, $u_l=0$ is the known equilibrium concentration and Eq.~\eqref{eq:solute} gives the evolution of the liquid fraction inside the envelope. The concentration of the solid growing inside the envelope at constant $\Omega$ is $u_s=1$.

The supersaturation across the stagnant film, $u_\delta$, needed to calculate the envelope growth velocity (Eq.~\eqref{eq:cantorvogel} and additional relations), is obtained from the concentration field in the liquid around the envelope, which is fully resolved numerically (Eq.~\eqref{eq:solute}). This means that the locally valid Ivantsov solution for diffusion around a dendrite tip is matched to the solution of the mesoscopically valid solute transport equation at a distance $\delta$ from the envelope (Figure~\ref{fig:schematics}c).

The solute conservation equation~\eqref{eq:solute} and the equation of the evolution of the indicator field used for envelope tracking are solved by a finite volume solver with the OpenFOAM\textsuperscript{\textregistered} toolbox. Second and higher order discretization schemes are used for the spatial operators. Implicit Euler time-stepping is used for the solute conservation equations and explicit time-stepping for the envelope indicator field equation. Adaptive time-stepping is employed and is useful during fast transients, such as the initial transient after nucleation. During other growth stages the time step is essentially limited by the characteristic diffusion time, $\Delta x^2/D$, at the scale of the mesh cell, $\Delta x$. In the present work uniform meshes with cubic cells are used. The mesh size needs to be adapted for an accurate description of the diffusion layer adjacent to the envelope, i.e. $\Delta x \lesssim 0.2 D/V$, where $V$ is a typical primary tip speed. A special search and interpolation algorithm is designed to provide the concentration at the distance $\delta$ from the envelope, needed for the calculation of the envelope velocity~\cite{Souhar_CMS_2016}. 

\section{Benchmarks}
\label{sec:benchmark}

Here, we compare the predictions of the three methods described above with a similar benchmark test case for undercooled isothermal equiaxed growth of a binary alloy. 
The phase field simulation uses well-converged numerical parameters, i.e. small enough grid spacing and diffuse interface width $W$, such that the PF results will be used as a reference for the discussion of the effect and choice of numerical parameters in the other two methods.

We set a solid seed (about as small as the method allows) in the corner of a cubic domain initialized at a given solute supersaturation $\Omega$. 
Thus, we simulate the growth of 1/8 of an equiaxed grain, using no-flux (mirror symmetry) Neumann boundary conditions on every face of the cube. 
We choose a value of the domain size $L$ (edge length of the cube) that is large enough to reach (or closely approach) a steady growth velocity, but also small enough to make simulations computationally achievable within a few days (less than two weeks for the longest) using quantitative phase-field using the current single-GPU-parallelized implementation.
Hence, we selected $L/l_D\approx2$, 4, 6, 8, and 10 for $\Omega=0.05$, 0.10, 0.15, 0.20, and 0.25, respectively.

In the PF simulations, we use a small interface energy anisotropy $\epsilon_4=0.007$ and a solute partition coefficient $k=0.1$.
In the DNN and GEM simulation we consider a tip selection constant $\sigma\approx0.0264$. 
This value was estimated from preliminary PF simulations of isothermal growth for $0.05\leq\Omega\leq0.25$ in a domain with $512^3$ grid points with tip velocities and radii measured as soon as the diffusion field gets affected at the side of the domain opposite to the dendrite. This value agrees with results from joint directional solidification simulation with similar $\epsilon_4$ and $k$. We noted a typical uncertainty of the order of 10 to 15\% depending upon the procedure chosen to extract the dendrite tip radius \cite{karma2000three}. Here, we follow the method described in reference \cite{clarke2017microstructure} (figure 5 therein), i.e. by 2D cross-section least-square fitting to a second order parabola over a fitting range that spans a distance of one tip radius behind the tip. A deeper analysis in terms of selection parameter is underway, such that the value used here only provides a reasonable estimation of $\sigma$ to use consistently in DNN and GEM simulations.

This value was estimated from independent directional solidification PF simulations with similar $\epsilon_4$ and $k$ parameters.
The selection parameter $\sigma$ estimated from the present isothermal PF simulations falls within 10\% of this value. However, directional solidification simulations reach a sustained steady-state even at lower undercooling, unlike the current simulations (see Figure~\ref{fig:comparison} discussed in section~\ref{sec:transient}).

The objective is to compare predicted growth velocities, not only in the steady-state, which should closely match the Ivantsov solution, but also and most importantly the initial transient, as the velocity tends towards the free dendrite steady-state, and the final transient, as the velocity declines to zero when the dendrite approaches the no-flux boundary wall.

For a given solute supersaturation $\Omega$, the theoretical P\'eclet number $P_s=\rho_s V_s/(2D)$ at steady-state (hence marked with a subscript ``$s$'')
is given by the Ivantsov solution~\cite{Ivantsov1947a}. 
It directly gives the ratio between the steady-state diffusion length $D/V_s$ and dendrite tip radius $\rho_s$. 
In this paper, most results are presented scaled with respect to the steady-state tip radius $\rho_s$ and velocity $V_s$.
For each value of $\Omega$ considered here, Table~\ref{tab:steady} summarizes the correspondence between these theoretical steady-state values and material parameters in physical units, such as the liquid solute diffusivity $D$ and the interface capillarity length $d_0$.
Within the selected supersaturation range, there is an order of magnitude of variation in the theoretical steady-state values of the P\'eclet and tip radius, and two orders of magnitude of variation of the tip velocity.

In the phase-field simulations, the finite difference grid element size is set equal to the diffuse interface width with $\Delta x=W\approx\rho_s/10$, which was found to yield well-resolved converged results.
The initial spherical seed has a radius $r_0=2.5\Delta x$, with a preconditioned phase field $\psi=r-r_0$ with $r=\sqrt{x^2+y^2+z^2}$ and a solute field $U=-\Omega$, which is equivalent to setting $c=c_\infty$ (i.e. $u=\Omega$) in the liquid and $c=kc_\infty$ in the solid.

In the DNN simulations, we initialize a solid seed with $u=0$ in the corner of the domain and setting $u=\Omega$ in the liquid. 
The seed consists of three quarters of paraboloids along the three normal grid directions with initial tip velocity $V=0$, radius $\rho=\rho_s$, and length $L_0$ larger than $r_{i}$ and $r_{max}$.
In particular, we study the effect of mesh spacing, $\Delta x$, and truncation radius, $r_{max}$. 
During the final transient, the integration sphere may be truncated if it exceeds the limits of the domain.

In the GEM simulations, we initialize a spherical envelope seed with $u=0$ and $g_l=1-\Omega$ in the corner of the domain and set $u=\Omega$ in the liquid. The size of the envelope seed is given by the smallest radius of curvature that can be accurately represented by the envelope tracking method on a mesh with spacing $\Delta x$~\cite{Sun2007,Souhar_CMS_2016}: $r_0=4.46 \Delta x$. The mesh spacing and time step needed for well-resolved converged results were determined by systematic convergence studies. Remaining mesh dependence, due to the initial seed size, which can be reduced on finer grids, is perceptible only during the initial growth transient and is discussed in Section~\ref{sec:resu:env}.

\begin{center}
\begin{table}[t]
\caption{
Theoretical steady-state P\'eclet number $P_s$, tip radius $\rho_s$, and velocity $V_s$, for different $\Omega$.
\label{tab:steady}
}
\centering
\begin{tabular}{c c c c}
\hline
\vspace{2pt}
$\Omega$ & $P_s=\frac{\rho_sV_s}{2D}$ & $\frac{\rho_s}{d_0} = \frac{1}{\sigma P_S}$ & $\frac{V_sd_0}{D} = 2\sigma P_s^2$\\
\hline
0.05 & 0.0131 & {2895} & {$9.03\cdot 10^{-6}$} \\
0.10 & 0.0341 & {1111} & {$6.14\cdot 10^{-5}$} \\
0.15 & 0.0624   & {607} & {$2.06\cdot 10^{-4}$} \\
0.20 & 0.0988   & {383} & {$5.15\cdot 10^{-4}$} \\
0.25 & 0.1446   & {261} & {$1.10\cdot 10^{-3}$} \\
\hline
\end{tabular}
\end{table}
\end{center}

\section{Results and Discussions}

\begin{center}
\begin{table}[!b]
\caption{Model parameters used with the three methods. 
}
\label{tab:model_params}
\centering
\begin{tabular}{c c c c c c c}
\hline
  & PF & \multicolumn{2}{c}{DNN} & GEM  \\
$\Omega$ & $W/d_0$ & $r_i/\rho_s$ & $r_{max}/\rho_s$ &  $\delta/(D/V_s)$\\
\hline
0.05 & 289 & 2.6 & 3,\,10 & {0.06} \\
0.10 & 110 & 2.4 & 3 & {0.20} \\
0.15 & 60 & 1.6 & 3 & {0.40} \\
0.20 & 38 & 1.9 & 3 & {0.60} \\
0.25 & 26 & 0.6 & 2 & {0.80} \\
\hline
\end{tabular}
\end{table}
\end{center}

Model parameters used for these simulations are summarized in Table~\ref{tab:model_params}. 
In order to provide a consistent illustration of the computational cost for all three methods, we also summarize the numerical parameters, namely the grid element size and time step for each $\Omega$ in Table~\ref{tab:num_params}.

The results presented here were obtained using model parameters that gave a good fit to the time evolution of the growth velocities obtained from PF simulations, while conserving a substantially lower computational cost. 
Since we used a wide range of computers (e.g. CPU/GPU) and methods (e.g. explicit/implicit time stepping), we did not perform a systematic performance analysis and focused on the potential of each method to yield quantitative predictions. Yet, in the following subsection, we provide values used for grid spacing and time steps, which provide a consistent picture of the relative computational workload to each method

\subsection{Steady-state growth}

\begin{figure}[!b]
\centering
\includegraphics[width=.775\columnwidth]{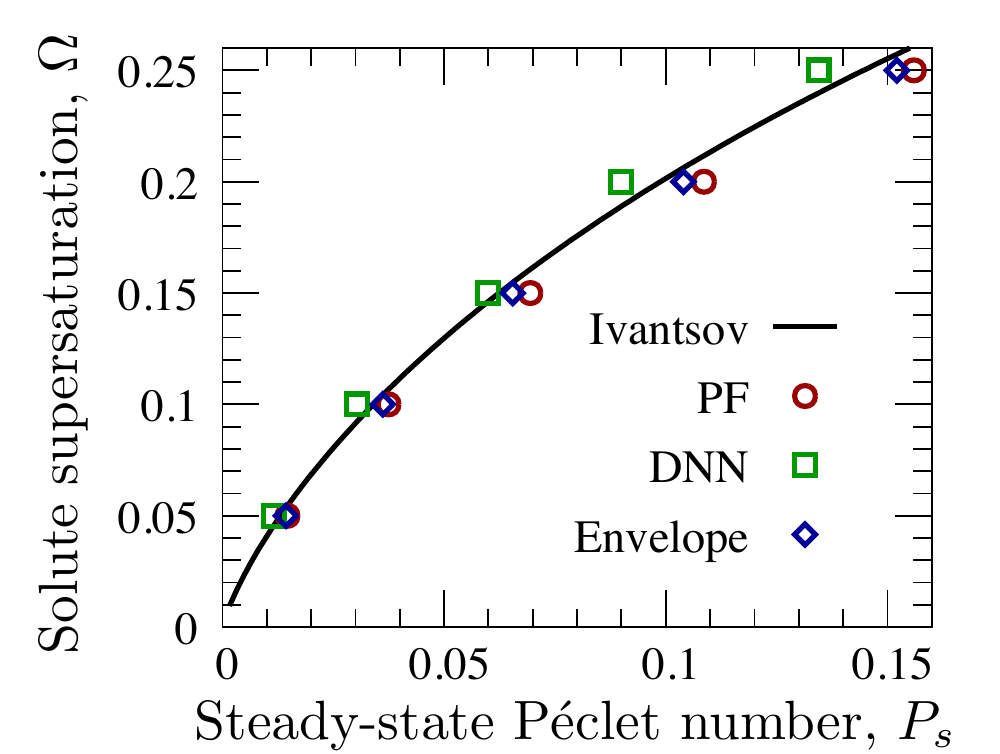}
\caption{
Predicted steady-state P\'eclet number for different imposed supersaturation $\Omega$ compared to the analytical Ivantsov solution~\cite{Ivantsov1947a}.
}
\label{fig:ivantsov}
\end{figure}

First, we compare steady-state growth predictions by all models for the range of solute supersaturation $0.05\leq\Omega\leq0.25$. 
Steady-state growth P\'eclet numbers for $\Omega=0.05$ to 0.25, predicted by PF, DNN, and GEM models, are compared to the theoretical Ivantsov solution in Figure~\ref{fig:ivantsov}. 
Values reported in figure 2 are averaged in time during steady-state growth, in order to integrate numerical oscillations. Such oscillation may occur as a dendrite tip crosses successive grid points and are due to the coarse grids used in DNN and GEM. 
The fluctuations in tip velocity and radius (see e.g. Fig.~\ref{fig:comparison} discussed later) lead to a negligible fluctuation in P\'eclet number (within the symbol size used in Fig.~\ref{fig:ivantsov}). 
In summary, all  models are capable of solving numerically the Ivantsov problem, and they thus appropriately predict the steady growth state.

\begin{center}
\begin{table*}[!t]
\caption{
Normalized mesh size and time step used with the three methods.
}
\label{tab:num_params}
\centering
\begin{tabular}{c c r r r c r r r}
\hline
$\Omega$	& & \multicolumn{3}{c}{$\Delta x/(D/V_s)$} &  &  \multicolumn{3}{c}{$\Delta t/(D/V_s^2)$} \\
			& & PF & DNN & GEM & & PF & DNN & GEM\\
\hline
0.05 & & {0.0026} & 0.0065 & {0.02} 	& &	{$10^{-6}$}	& {$7\cdot10^{-6}$}	&	{$10^{-4}$}\\
0.10 & & {0.0068} & 0.017 & {0.04} 	& &	{$7\cdot10^{-6}$}	& {$4\cdot10^{-5}$}	&	{$4\cdot10^{-4}$}\\
0.15 & & {0.0125} & 0.031 & {0.06} 	& &	{$2\cdot10^{-5}$}	& {$10^{-4}$}	&	{$9\cdot10^{-4}$}\\
0.20 & & {0.0198} & 0.049 & {0.08} 	& &	{$6\cdot10^{-5}$}	& {$4\cdot10^{-4}$}	&	{$10^{-3}$}\\
0.25 & & {0.0289} & 0.072 & {0.10} 	& &	{$10^{-4}$}	& {$8\cdot10^{-4}$}	&	{$2\cdot 10^{-3}$}\\
\hline
\end{tabular}
\end{table*}
\end{center}

\begin{figure*}[!t]
\centering
 \includegraphics[width=.8\textwidth]{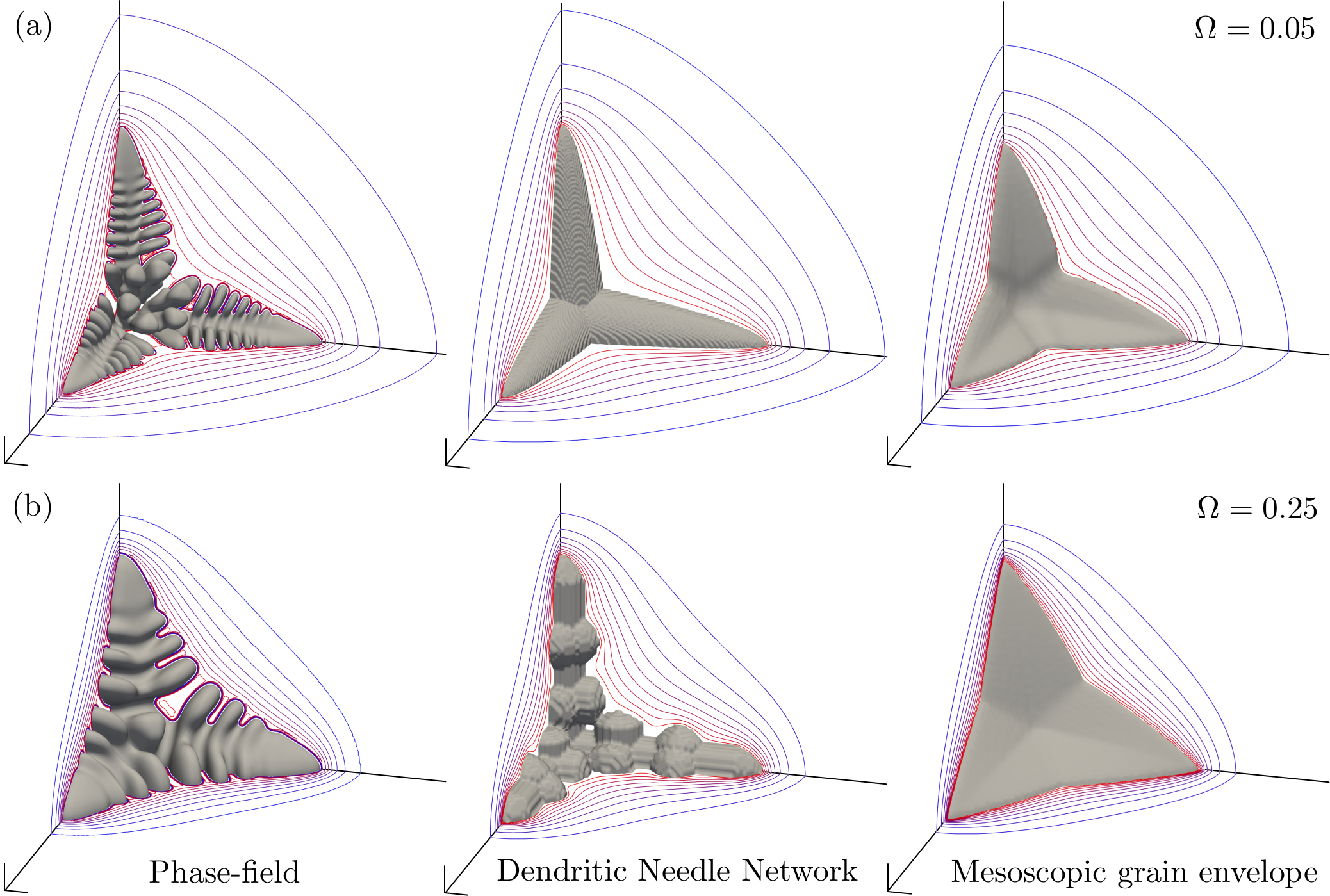}
\caption{
Grain shape (gray) and iso-concentration lines,
 from $u=0$ (red) to $u=0.9\Omega$ (blue) with steps of $0.1\Omega$, 
 along the ($x=0$), ($y=0$), and ($z=0$) planes when the length of the primary dendritic trunk reaches one half of the simulation domain for $\Omega=0.05$ (a) and 0.25 (b) simulated using phase-field (left), DNN (middle), and Envelope (right) methods.
The total size of the domain, illustrated by the visible bottom left corner of each simulation, is $2D/V_s\approx 221\,400\,d_0$ for $\Omega=0.05$ and $10D/V_s\approx9\,060\,d_0$ for $\Omega=0.25$.
}
\label{fig:shapes}

\end{figure*}

\begin{figure*}[!t]
\centering
\includegraphics[width=.775\textwidth]{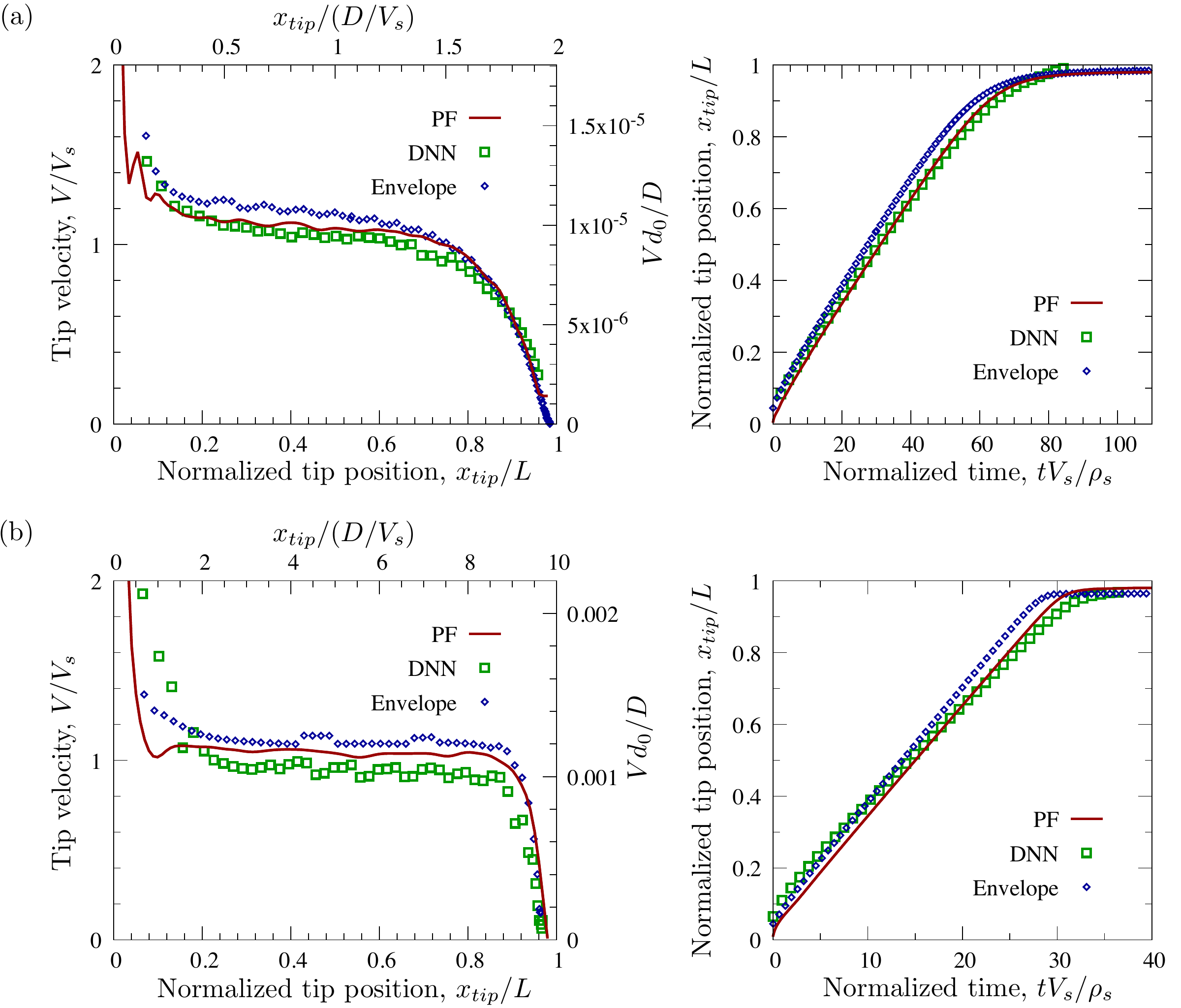} 
\caption{
Evolution of the primary dendrite tip velocity from the initial to the final transient state (left) and primary trunk length versus time (right) for $\Omega=0.05$ (a) and $\Omega=0.25$ (b).
In order to allow comparing models all together at different supersaturations, time and space are scaled with respect to the theoretical steady-state velocity $V_s$ and tip radius $\rho_s$. 
The velocity evolution is plot as a function of the tip location normalized by the domain length.
Note that left-hand-side velocity plots have been smoothed to reduce numerical oscillations and improve readability (right-hand-side length plots are raw).
}
\label{fig:comparison}
\end{figure*}

The shape of the predicted grain when the length of the primary trunk reaches one half of the domain size is shown in Figure~\ref{fig:shapes} for all three models for $\Omega=0.05$ (top) and $\Omega=0.25$ (bottom).
For the sake of illustration, the DNN microstructure is shown for a simulation performed without sidebranches for $\Omega=0.05$ and no thickness bounding (i.e. with $r_{max}=\infty$), and with sidebranches and $r_{max}=2\rho_s$ for $\Omega=0.25$.
In the latter case, the effect of a finite $r_{max}$ becomes apparent from the cylindrical shape of the branches. 
While at this stage the velocity has already started decreasing due to the effect of boundaries, the shape of the microstructures is barely distinguishable from that at steady-state, for all three models.


\subsection{Transient growth}
\label{sec:transient}

Next, we compare the predicted transient evolution of grain growth velocities. We focus on $\Omega=0.05$ and $\Omega=0.25$, which are illustrative of results obtained at intermediate $0.05<\Omega<0.25$. 
The evolution of the velocity from initial to final transient appears in Figure~\ref{fig:comparison} for $\Omega=0.05$ (a) and $
\Omega=0.25$ (b).
The velocity is shown as a function of the tip location normalized by the domain length.
This way, we filter out the possible different onset times for the final decrease of the velocity --- such differences arise when the steady-state is maintained at a different velocity.
The right-hand-side panels, showing the evolution of the primary trunk length versus time, illustrate that the oscillations appearing in the left-hand-side velocity plot have a negligible effect on the length of the dendrite, once integrated over time.
These results show that, for appropriately chosen sets of parameters, both DNN and GEM models can reproduce PF results with reasonable accuracy.

DNN simulations were performed with $\Delta x=0.25\rho_s$, $r_i=2.6\rho_s$, $r_{max}=3\rho_s$ for $\Omega=0.05$, and with $\Delta x=0.25\rho_s$, $r_i=0.6\rho_s$, $r_{max}=2\rho_s$ for $\Omega=0.25$ (see Table~\ref{tab:model_params}).
We chose a resolution providing a compromise between accuracy and simulation time. 
The chosen grid size is smaller than the typical value used in DNN simulations, i.e. $\Delta x \approx \rho_s$ or higher. 
Other parameters are close to typical parameters used in the DNN model, i.e. fulfilling $r_i\ll l_D$ and $r_{max}$ chosen as to not bound the needle thickness within the flux integration domain.
Interestingly, we notice very little influence of sidebranching on the velocity of the primary tip, both in steady and transient states.
It is also worth noting that the integration radius $r_i$ can be taken lower than the tip radius, as long as the integration domain contains a sufficient number of integration points.

GEM simulations were performed with a stagnant film thickness of $\delta=0.06 D/V_s$ and $0.80 D/V_s$, and with a mesh spacing of $\Delta x=0.02 D/V_s$ and $0.10 D/V_s$, for $\Omega=0.05$, and 0.25, respectively. The $\delta$ and the mesh used at high $\Omega$ are close to those already recommended by Souhar et al.~\cite{Souhar_CMS_2016} for steady-state equiaxed growth. At low $\Omega$, significantly smaller $\delta$ were needed to accurately match both the initial and the final transient of the tip evolution. The same conclusion was recently made by Olmedilla et al.~\cite{Olmedilla_AM_2019} when comparing the GEM to experiments on transient equiaxed growth at comparably low undercoolings. The mesh spacing is dictated by the diffusion length at high $\Omega$ ($\Delta x\approx0.1 D/V_s$ here), however at low $\Omega$ the small stagnant-film thickness, $\delta$, becomes the limiting factor. For an accurate calculation and interpolation of $u_\delta$ the mesh spacing should be $\Delta x<0.5 \delta$. 

\subsection{Effect of DNN parameters}

\begin{figure}[t]
\centering
\includegraphics[width=.81\columnwidth]{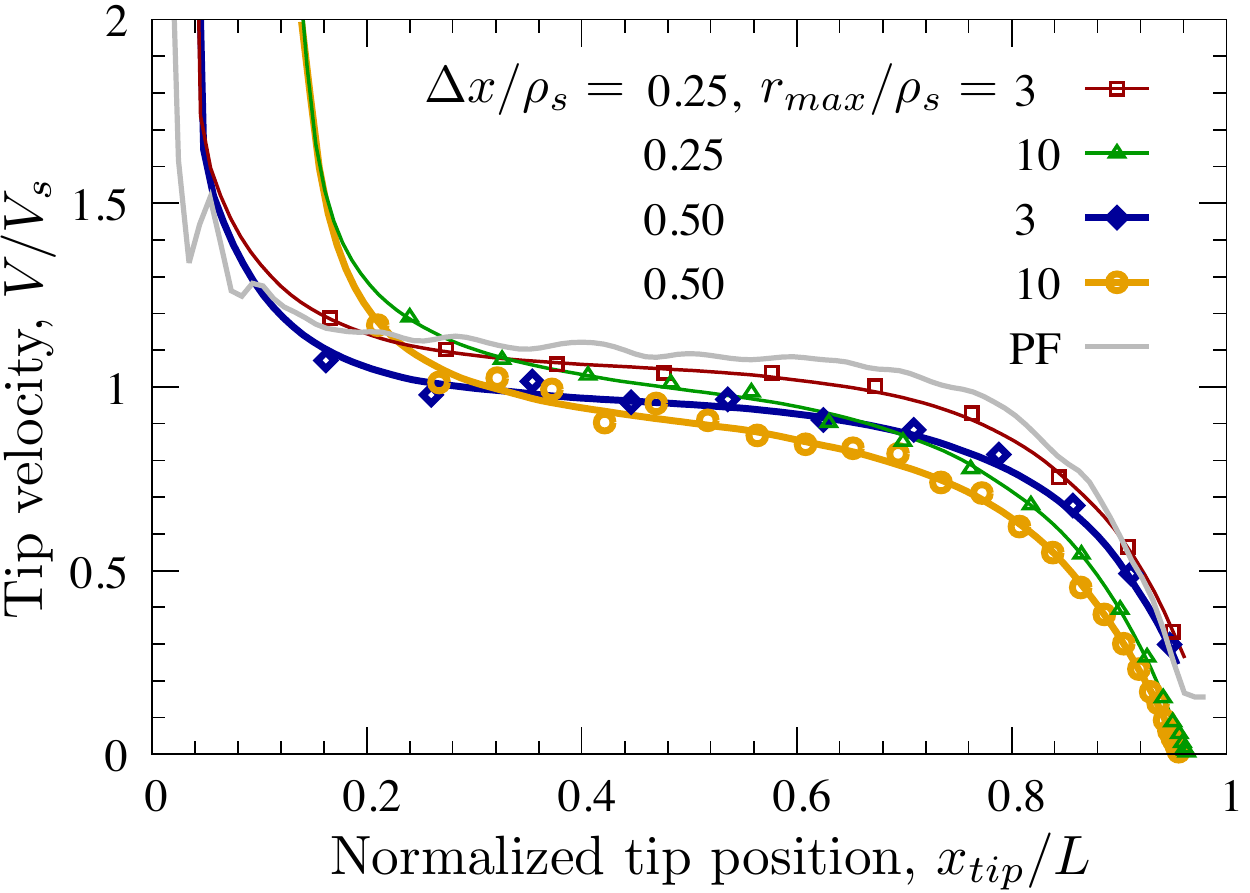}
\caption{
\label{label}
Evolution of the primary dendrite tip velocity for $\Omega=0.05$ from DNN simulations, showing the effect of grid spacing $\Delta x$ and stem truncation radius $r_{max}$.
In all runs, the radius of the integration sphere $r_i/\rho_s$ was kept to 2.6 or 2.7. 
For readability, symbols show raw data whereas lines show smoothed data (filtering numerical oscillations as the tip progresses between successive grid points).
}
\label{fig:dnn}
\end{figure}

Figure~\ref{fig:dnn} shows the tip velocity evolution predicted by DNN simulations for different combinations of grid spacing $\Delta x$ and needle truncation radius $r_{max}$, for a given flux integration radius $r_i=2.6\sim2.7\rho_s$ and $\Omega=0.05$.

Decreasing the grid spacing $\Delta x$ for identical values of $r_{max}$ and $r_i$ results mainly in (i) less fluctuation of the velocity and (ii) higher velocities, in better agreement with PF results. 
On the other hand, the effect of increasing the radius of the stem truncation $r_{max}$ for identical values of $\Delta x$ and $r_i$ is to decrease the steady-state tip velocity.

The effect of the initial length of the seed branches needs to be discussed. The initial length $L_0$ of the branches differ among the simulations because the simulations are initialized with paraboloids of length $L_0=r_{max}+\Delta x$. 
The difference in initial length most notably affects the growth kinetics during the initial and final transient stages.
All four simulations nonetheless stabilize close to the theoretical steady-state velocity given by the Ivantsov solution.
This plateau region, however, lasts for a shorter time for the simulations with larger $r_{max}$ and hence longer $L_0$.
At this low $\Omega$, in order for PF simulations to be achievable within a few days, we limited the domain size to 2$l_D$.
For this reason, the amount of time spent in steady-state is limited, and for high $L_0$ values, the final transient sets in before steady-state is fully maintained.
In the last stage of the final transient, a velocity plateau may be observed (not represented here but visible by interrupted plots) due to a numerical restriction set to the tip velocity when the distance to the wall becomes lower than $r_i$.
This limitation might be accommodated by either (i) truncating the integration domain, (ii) integrating boundary conditions within the integration of $\cal F$, or (iii) using an adaptive $r_i$ that decreases when approaching a boundary.

\subsection{Effect of Envelope parameters}
\label{sec:resu:env}

\begin{figure*}[b]
\centering
\includegraphics[width=.81\textwidth]{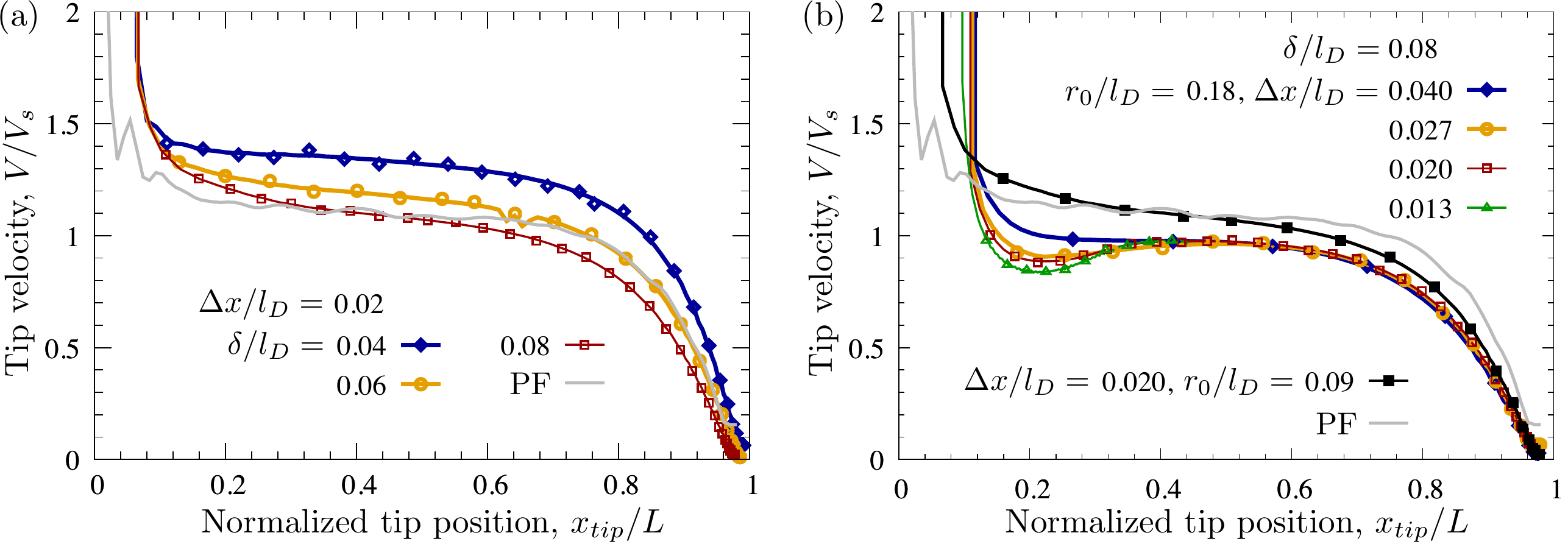}
\caption{
Evolution of the primary dendrite tip velocity for $\Omega=0.05$ from GEM simulations, showing the effect of the model and numerical parameters. (a) The effect of the stagnant film thickness, $\delta$; all simulations with a mesh $\Delta x/(D/V_s)=0.02$. (b) The effect of the mesh spacing $\Delta x$, and of the size of the initial nucleus, $r_0$; all simulations with $\delta/(D/V_s)=0.08$.
}
\label{fig:env}
\end{figure*}

The main model parameter of the GEM is the stagnant film thickness, $\delta$. Table~\ref{tab:model_params} summarizes the $\delta$ that gave the best match to the tip evolution predicted by phase field. The best $\delta$ clearly increases with $\Omega$. The dependence of the transient growth velocity on $\delta$ at $\Omega=0.05$ is shown in {Figure~\ref{fig:env}a} for a fixed stagnant film thickness $\delta/(D/V_s)=0.08$. The most striking observation is the effect on the steady-state velocity (already pointed out by Souhar et al.~\cite{Souhar_CMS_2016}), which affects the whole evolution. The de- and accelerations during the initial and final transient do not significantly depend on $\delta$.

{Figure~\ref{fig:env}b} shows the influence of the mesh spacing on the transient tip growth at $\Omega=0.05$. The influence of the mesh spacing is twofold. In addition to its direct impact on the solution of the PDEs, $\Delta x$ also limits the smallest radius of the initial nucleus, $r_0$, and thus the initial conditions (see Section~\ref{sec:benchmark}). Typically, $r_0 \geq 4.46 \Delta x$. The direct impact of $\Delta x$, \emph{excluding} that of $r_0$, is shown by four simulations in which $r_0$ was kept fixed and only the mesh spacing was varied from $0.013 D/V_s$ to $0.04 D/V_s$. The initial radius was $r_0 = 0.18 D/V_s$ which is the smallest $r_0$ that can be resolved on the coarsest of these four grids. The differences between the four solutions are strictly limited to the initial transient, and perfectly overlap at larger arm lengths. The reason for this is that the thin diffusion layer around the grain during the initial stages of growth requires a fine mesh resolution to be accurately resolved. Once the tip decelerates and the diffusion layer grows wider, coarse meshes are sufficient. 

However, the most important influence related to the mesh resolution is actually that of the initial nucleus size. This is shown through a comparison of the evolution of tip velocity for $r_0=0.18 D/V_s$ and $0.09 D/V_s$ with the same $\Delta x=0.02D/V_s$ (open versus full square symbols in Figure~\ref{fig:env}b). We note a much slower deceleration during the initial transient at smaller $r_0$. Furthermore, the steady growth stage is replaced by a smooth transition to the final transient, which is not significantly affected.

In summary, accurate simulation of transient equiaxed growth requires a stagnant film thickness that roughly scales as $\delta \sim \Omega D/V_s$. This does not contradict, but goes beyond prior results~\cite{Souhar_CMS_2016} that established a recommendation for \emph{steady-state} growth: $\delta \sim D/V_s$. If compared, the new recommendation is slightly less accurate in steady-state (with a difference of $<10\%$) but is more accurate during transients. 

The mesh spacing and the initial nucleus radius have a significant influence on the results only during the initial growth transient, but not during steady-state and the final transient. Note that because of the relatively small domain size ($L=2 D/V_s$) in the case presented here ($\Omega=0.05$), the impact of the initial transient on the whole evolution is prominent. This is not the case at higher $\Omega$, where larger domains are used and the initial transient is a smaller portion of the growth. Generally, $\Delta x \sim 0.1 D/V_s$ is sufficiently fine for steady-state and for the final growth transient, but should be refined if good accuracy during initial transients is required. Furthermore, $\Delta x < 0.5 \delta$ is an additional constraint that is important at low $\Omega$, where smaller $\delta/(D/V_s)$ must be used.
 
\section{Summary and Perspectives}

We have compared the results of phase-field, dendritic needle network, and grain envelope models for a simulation of the equiaxed growth of an isothermally undercooled binary alloy.
Using well-converged PF results as the benchmark reference, we discussed the effect and choice of model and numerical parameters in the DNN and GEM methods upon their accurate prediction of steady and transient dendrite growth velocities.

As expected, all models can appropriately predict steady-state growth conditions prescribed by microscopic solvability and Ivantsov solutions.
Moreover, in the transient regimes, both GEM and DNN can accurately describe growth velocities, if appropriate model and numerical parameters are selected. 
The final transients (solutal interactions between grains in practical applications) are typically accurate with standard parameters.
The early-stage growth transient, on the other hand, requires a fine tuning of model parameters in order to achieve quantitative predictions, such that the initial seed size and mesh spacing need to be finer than typical values recommended in previous works, in both DNN and GEM approaches. 

While the present study provides already some new insight and guidance for the use of these relatively new modeling techniques, these results remain preliminary, and the study is ongoing.
Work in progress on the current benchmarks include: 
(i) the extraction of scaling laws for initial and final transient growth stages,
(ii) further systematic exploration of model parameters (e.g. $r_i$ for DNN),
(iii) comparison of envelope shapes in equiaxed growth, which is important in the scope of grain interactions, 
and
(iv) exploring the advantages and limitations of models at higher supersaturations.
Perspectives for future benchmarks include: 
(i) benchmarks for directional solidification, in particular in the scope of spacing selection, 
and 
(ii) configurations with convection in the liquid phase.

From the results of these studies, we expect that new numerical techniques will improve the predictive capabilities of the mesoscopic models.
Such techniques could include, for instance, the introduction of adaptive parameters to ensure better treatment of transient regimes, e.g. using adaptive $r_i$ in DNN or $\delta$ in GEM that could depend upon the growth acceleration/deceleration.
We also expect that the present work, and those to follow, will provide useful guidance to potential users and developers of these promising tools for bridging length and time scales in the modeling of dendritic solidification.

\section*{Acknowledgements}
D.T. acknowledges support from the European Union's Horizon 2020 research and innovation programme under a Marie Sk\l odowska-Curie Individual Fellowship (Grant Agreement 842795).
L.S. and A.V. acknowledge support by the German Ministry of Economy via the Deutsche Zentrum f\"ur Luft- und Raumfahrt, DLR (Contract 50WM1743).
MZ acknowledges support by the French State through the program ``Investment in the future'' operated by the National Research Agency (ANR) and referenced by ANR-11 LABX-0008-01 (LabEx DAMAS).


\bibliographystyle{unsrt}
\bibliography{2020_MCWASP_Benchmarks}

\end{document}